\documentclass[twocolumn,superscriptaddress,showpacs,prl]{revtex4}
\usepackage{graphicx}

\newcommand{\ket}[1]{|{#1}\rangle}

\begin{document}
\title{Faithful qubit distribution assisted by one additional qubit against collective noise }

\author{T. Yamamoto}
\email{yamamoto@mp.es.osaka-u.ac.jp}
\affiliation{Division of Materials Physics, Department of Materials Engineering Science,
Graduate school of Engineering Science, Osaka University, Toyonaka, Osaka 560-8531, Japan}
\affiliation{CREST Research Team for Photonic Quantum Information, 4-1-8 Honmachi, Kawaguchi, Saitama 331-0012, Japan}

\author{J. Shimamura}
\affiliation{Division of Materials Physics, Department of Materials Engineering Science, Graduate school of Engineering Science, Osaka University, Toyonaka, Osaka 560-8531, Japan}
\affiliation{CREST Research Team for Photonic Quantum Information, 4-1-8 Honmachi, Kawaguchi, Saitama 331-0012, Japan}
\affiliation{SORST Research Team for Interacting Carrier Electronics, 4-1-8 Honmachi, Kawaguchi, Saitama 331-0012, Japan}
\author{\c{S}. K. \"Ozdemir}
\affiliation{Division of Materials Physics, Department of Materials Engineering Science,
Graduate school of Engineering Science, Osaka University, Toyonaka, Osaka 560-8531, Japan}
\affiliation{CREST Research Team for Photonic Quantum Information, 4-1-8 Honmachi, Kawaguchi, Saitama 331-0012, Japan}
\affiliation{SORST Research Team for Interacting Carrier Electronics, 4-1-8 Honmachi, Kawaguchi, Saitama 331-0012, Japan}
\author{M. Koashi}
\affiliation{Division of Materials Physics, Department of Materials Engineering Science, Graduate school of Engineering Science, Osaka University, Toyonaka, Osaka 560-8531, Japan}
\affiliation{CREST Research Team for Photonic Quantum Information, 4-1-8 Honmachi, Kawaguchi, Saitama 331-0012, Japan}
\affiliation{SORST Research Team for Interacting Carrier Electronics, 4-1-8 Honmachi, Kawaguchi, Saitama 331-0012, Japan}
\author{N. Imoto}
\affiliation{Division of Materials Physics, Department of Materials Engineering Science,
Graduate school of Engineering Science, Osaka University, Toyonaka, Osaka 560-8531, Japan}
\affiliation{CREST Research Team for Photonic Quantum Information, 4-1-8 Honmachi, Kawaguchi, Saitama 331-0012, Japan}
\affiliation{SORST Research Team for Interacting Carrier Electronics, 4-1-8 Honmachi, Kawaguchi, Saitama 331-0012, Japan}

\date{\today}

\begin{abstract}
We propose a distribution scheme of polarization states of a single photon over collective-noise channel. By adding one extra photon with a fixed polarization, we can protect the state against collective noise via a parity-check measurement and post-selection. While the scheme succeeds only probabilistically, it is simpler and more flexible than the schemes utilizing decoherence-free subspace. An application to BB84 protocol through collective noise channel, which is robust to the Trojan horse attack, is also given. 
\pacs{03.67.Pp, 03.67.Dd, 03.67.Hk}
\end{abstract}

\maketitle

Protecting quantum information from decoherence process is essential to exploit the information processing power provided by quantum physics, such as quantum key distribution (QKD), quantum computation, and quantum teleportation.  In optical QKD protocols, especially, one of the main sources of noise is the fluctuation of the birefringence of the optical fiber which alters the polarization state of photons. Usually, the fluctuation 
is slow in time, so that the alteration of the polarization is considered to be the same over the sequence of several photons. Such noise is referred to as collective noise (or correlated noise). To overcome such noise, several elegant QKD schemes have been developed so far \cite{Gisin02}. One of these schemes is based on the phase difference of single photon in two sequential time bins, which is achieved by two unbalanced Mach-Zehnder interferometers(MZIs) shared between the sender and the receiver \cite{Bennett92,Townsend93}. Although this scheme achieves faithful QKD through optical fiber connecting two unbalanced MZIs, one needs active stabilization to adjust the phase differences of each unbalanced MZI with the precision of the order of the wavelength. To circumvent the active stabilization of the MZIs, the Plug-and-play system \cite{Muller97} has been proposed and developed for QKD. In this scheme, the fluctuation of optical fiber is passively compensated via a two-way quantum communication with the use of the Faraday orthoconjugation effect. However, the use of two-way quantum communication makes the system vulnerable to an eavesdropping technique known as Trojan horse attack, especially for the single photon based QKD. 

To avoid the eavesdropping by Trojan horse attack, several novel QKD schemes which rely on one way quantum communication have recently been proposed \cite{Wang03,Walton03,Boileau04,Boileau04qph}. 
In these schemes, one first encodes the quantum states into the proper multi-photon entangled states in the decoherence-free (DF) subspaces \cite{Lidar03}, then sequentially distributes each photon through the optical fiber. The states in the DF subspaces are invariant under collective noise, leading to faithful QKD through one-way quantum communication.

In this paper, we present a distribution scheme for the single photon polarization state by one-way quantum communication, which requires no encoding process. The present scheme is immune to collective noise and can be used for QKD.
Unlike other schemes\cite{Wang03,Walton03,Boileau04,Boileau04qph} for QKD against collective noises, it is accomplished just by sending an additional photon in a {\it fixed} polarization, and it is not necessary to encode a qubit into an entangled state of multiple qubits in DF subspace. Due to the lack of entanglement in the initially prepared state, the scheme succeeds only probabilistically. 
Instead, the simplicity of the present scheme will make it easy to apply for many other quantum communication protocols based on qubits. 
For example, if one of the parties in a multi-party protocol wants to send his qubit to another party in the protocol, faithful qubit distribution can be achieved by adding just an additional photon in a {\it fixed} polarization and applying the present scheme.

\begin{figure}[tbp]
\begin{center}
 \includegraphics[scale=0.7]{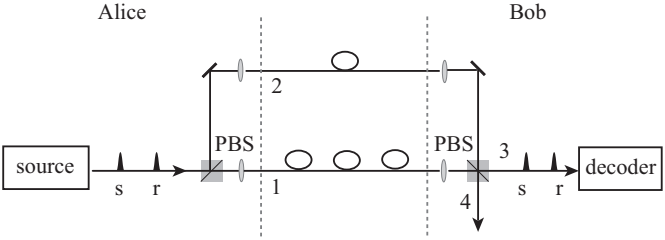}
\end{center}
 \caption {The schematic diagram of the proposed distribution scheme.
 The photons for the reference r and the signal s, which are temporally separated by $\Delta t$, are split into two spatial modes by a 
 PBS on Alice's side. These photons in  the polarizations $\ket{H}_{\rm r,s}$ and $\ket{V}_{\rm r,s}$ are distributed to Bob passing through the quantum channels 1 and 2, respectively. After reaching Bob, these photons are mixed by a PBS. Bob then extracts the signal state from the received two photons.
}
\label{fig:setup}
\end{figure}

Let us first introduce our distribution scheme and describe the possible implementation by a single-photon source, linear optics and photon detectors.
The sender, Alice, first prepares a photon in the state $\ket{D}_{\rm
r}\equiv 1/\sqrt{2}(\ket{H}_{\rm r}+\ket{V}_{\rm r})$ as a reference,
followed by  a signal photon in an arbitrary pure state $\alpha\ket{H}_{\rm s}+\beta\ket{V}_{\rm s}$ in the same spatial mode after a delay $\Delta t$. Here $\ket{H}$ and $\ket{V}$ represent horizontal and vertical polarization states, respectively.
The state of these two photons can be written as
\begin{eqnarray}
\ket{D}_{\rm r}\otimes(\alpha\ket{H}_{\rm s}+\beta\ket{V}_{\rm s}).
\label{eq:1}
\end{eqnarray}
As shown in Fig.~\ref{fig:setup}, these photons are split into two
spatial modes by a polarizing beamsplitter (PBS), which transmits
$\ket{H}$ and reflects $\ket{V}$. Alice then sends the photons in $\ket{H}_{\rm r,s}$ and $\ket{V}_{\rm r,s}$ through the
 channels 1 and 2, respectively.  We assume that each channel
is composed of a polarization maintaining optical fiber (PMF). Bob
receives these photons and  mixes them by a PBS. If there is no
fluctuation or loss in the fibers, one can adjust the optical
path lengths of the two fibers so that the state of the two
photons in mode 3 is unaltered from the state (\ref{eq:1}).

To see how we obtain the signal state against collective dephasing, suppose that  the channels 1 and  2 add unknown phase shifts $\phi_{H}$ and $\phi_{V}$, respectively, due to the fluctuations in the optical path length in PMF. We assume that these fluctuations are slower than the interval $\Delta t$, namely, $\phi_{H}(t)=\phi_{H}(t+\Delta t)=\phi_{H}$ and $\phi_{V}(t)=\phi_{V}(t+\Delta t)=\phi_{V}$. In this case, the states $\ket{H}_{\rm r,s}$ and $\ket{V}_{\rm r,s}$ are transformed into $e^{i\phi_{H}}\ket{H}_{\rm r,s}$ and $e^{i\phi_{V}}\ket{V}_{\rm r,s}$, respectively. Therefore, Bob receives the photons in the state
\begin{eqnarray}
& &1/\sqrt{2}[\alpha e^{2i\phi_{H}}\ket{H}_{\rm r}\ket{H}_{\rm s}+\beta e^{2i\phi_{V}}\ket{V}_{\rm r}\ket{V}_{\rm s} \nonumber \\
& &\qquad +e^{i(\phi_{H}+\phi_{V})}(\alpha \ket{V}_{\rm r}\ket{H}_{\rm s}+\beta \ket{H}_{\rm r}\ket{V}_{\rm s})].
\label{eq:3}
\end{eqnarray}
It is easy to see that the state $\alpha \ket{V}_{\rm r}\ket{H}_{\rm s}+\beta \ket{H}_{\rm r}\ket{V}_{\rm s}$ is invariant under these phase fluctuations. We could, in principle, project the state (\ref{eq:3}) onto  the state $\alpha \ket{V}_{\rm r}\ket{H}_{\rm s}+\beta \ket{H}_{\rm r}\ket{V}_{\rm s}$ with a probability of $1/2$. If the successful projection occurs, then we decode the states $\ket{V}_{\rm r}\ket{H}_{\rm s}$ and $\ket{H}_{\rm r}\ket{V}_{\rm s}$ into  $\ket{H}$ and $\ket{V}$, respectively. We would thus recover the state of the signal qubit $\alpha \ket{H}+\beta \ket{V}$.

\begin{figure}[tbp]
\begin{center}
 \includegraphics[scale=1]{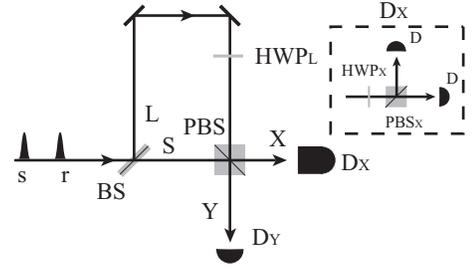}
\end{center}
 \caption {A decoder for the present scheme.
The received photons are split into two optical paths by a 
BS, then mixed by a PBS. HWP$_{\rm L}$ rotates the polarization by 90$^\circ$. The detector D$_{\rm X}$ consists of HWP$_{\rm X}$ which rotates the polarization by 45$^\circ$, a PBS, and photon detectors D as shown in the dashed box. The polarization state of the photon in the mode X is projected onto the diagonal basis  $\{ \ket{D}_{\rm X}, \ket{\bar{D}}_{\rm X} \}$.
The optical path difference between short path S and long path L is adjusted to compensate the temporal difference $\Delta t$ between the reference and signal photons.  The detector D$_{\rm Y}$ can be used to post-select the successful event by the twofold coincidence detection as described in the text. 
}
\label{fig:setupDecode}
\end{figure}

An implementation of the above projection and decoding by using linear optical elements and photon detectors is shown in Fig.~\ref{fig:setupDecode}. Bob first splits two photons in mode 3 into long path L and short path S by a polarization-insensitive beamsplitter (BS), then mixes these photons by a PBS. Half wave plate (HWP$_{\rm L}$) rotates the polarization of the photons in the path L by 90$^\circ$. The optical path difference between the paths L and S is adjusted to compensate the temporal difference $\Delta t$ between the reference and signal photons. The polarization analyzer D$_{\rm X}$ consists of a HWP$_{\rm X}$ rotating the polarization by 45$^\circ$, a PBS$_{\rm X}$, and two photon detectors. Using this analyzer, Bob projects the polarization of the photon in the mode X onto the diagonal basis $\{ \ket{D}_{\rm X}, \ket{\bar{D}}_{\rm X} \}$, where $\ket{\bar{D}}_{\rm X}\equiv 1/\sqrt{2}(\ket{H}_{\rm X}- \ket{V}_{\rm X})$.

Let us first consider the case where the reference photon passes through
the path L, and the signal photon through the path S. In this case, these photons reach the PBS at the same time.
As in the parity checking described in Ref.\cite{Pittman01,TYamamoto03,Pan03},
the incoming photons in the states $\ket{H}_{\rm r}\ket{H}_{\rm s}$ and
$\ket{V}_{\rm r}\ket{V}_{\rm s}$
(the two photons have different polarizations just before the PBS)
result in a two-photon state  and the
vacuum in the mode X, respectively. On the other hand, the photons in the 
state $\alpha \ket{V}_{\rm r}\ket{H}_{\rm s}+\beta \ket{H}_{\rm
r}\ket{V}_{\rm s}$ (the two photons have the same  polarization just before the PBS)
result in one photon in the mode X and one photon in Y, in the state $\alpha \ket{H}_{\rm Y}\ket{H}_{\rm X}+\beta \ket{V}_{\rm Y}\ket{V}_{\rm X}$. Hence, by counting the number of photons in the mode X, we can discard the cases $\ket{H}_{\rm r}\ket{H}_{\rm s}$ and $\ket{V}_{\rm r}\ket{V}_{\rm s}$.
If Bob detects a photon in  $\ket{D}_{\rm X}$,
the two photons are projected onto the state $\ket{D}_{\rm X}(\alpha
\ket{H}_{\rm Y}+\beta \ket{V}_{\rm Y})$. Then Bob obtains the photon in the 
 signal state $\alpha \ket{H}_{\rm Y}+\beta \ket{V}_{\rm Y}$. If Bob
detects a photon in  $\ket{\bar{D}}_{\rm X} $, two photons are
projected onto the state $\ket{\bar{D}}_{\rm X}(\alpha \ket{H}_{\rm
Y}-\beta \ket{V}_{\rm Y})$. In this case, Bob simply adds the phase
shift $\pi$ between  $\ket{H}_{\rm Y}$ and $\ket{V}_{\rm Y}$
to obtain  the photon in the signal state $\alpha \ket{H}_{\rm Y}+\beta \ket{V}_{\rm Y}$.

In a practical communication task such as QKD, photons in mode Y may be measured on a polarization basis. In this case, one can post-select the event where at least one photon is emitted in each mode X and Y by the twofold coincidence detection. Therefore conventional threshold photon detectors, which do not discriminate two or more photons from a single photon, can be used.

In the other cases where the reference photon passes through the path S
and/or the signal photon passes through the path L, at least one of the
photons should have a different arrival time at detector
D$_{\rm X}$ or D$_{\rm Y}$.
Hence we can eliminate the contributions from such cases
by discriminating the arrival times of the photons at the detectors
D$_{\rm X}$ and D$_{\rm Y}$ by post-selection. Any loss in the channels
and inefficiency of the detectors are also discarded by
the post-selection.

So far, we have described the immunity of the present scheme from
collective dephasing, which is the main source of errors when we use
 PMFs for the channels.
 In the following, we show that the present scheme also works against
 any collective polarization rotation (any collective unitary on the
qubit space) which occurs in two quantum
 channels independently.
Suppose that each collective polarization rotation transforms
the polarization states as
$\ket{H}_{\rm r,s}\to  \delta_1 \ket{H}_{\rm r,s}+\gamma_1 \ket{V}_{\rm r,s}$ and
$\ket{V}_{\rm r,s}\to  \delta_2 \ket{H}_{\rm r,s}+\gamma_2 \ket{V}_{\rm r,s}$ in quantum channel 1 and 2, respectively, where $|\delta_1|^2+|\gamma_1|^2=1$ and $|\delta_2|^2+|\gamma_2|^2=1$.
These rotations transform the state (\ref{eq:1}) into
\begin{widetext}
\begin{eqnarray}
1/\sqrt{2} [ \alpha (\delta^2_1\ket{H}_{\rm r3}\ket{H}_{\rm s3}+\delta_1\gamma_1\ket{H}_{\rm r3}\ket{V}_{\rm s4}
+\delta_1\gamma_1\ket{V}_{\rm r4}\ket{H}_{\rm s3}+\gamma^2_1\ket{V}_{\rm r4}\ket{V}_{\rm s4}) \nonumber \\
+\beta (\delta^2_2\ket{H}_{\rm r4}\ket{H}_{\rm s4}+\delta_2\gamma_2\ket{H}_{\rm r4}\ket{V}_{\rm s3}
+\delta_2\gamma_2\ket{V}_{\rm r3}\ket{H}_{\rm s4}+\gamma^2_2\ket{V}_{\rm r3}\ket{V}_{\rm s3}) \nonumber \\
+\alpha (\delta_1\delta_2\ket{H}_{\rm r4}\ket{H}_{\rm s3}+\gamma_1\delta_2\ket{H}_{\rm r4} \ket{V}_{\rm s4}
+\delta_1\gamma_2\ket{V}_{\rm r3}\ket{H}_{\rm s3}+\gamma_1\gamma_2\ket{V}_{\rm r3}\ket{V}_{\rm s4}) \nonumber \\
+\beta (\delta_1\delta_2\ket{H}_{\rm r3}\ket{H}_{\rm s4}+\delta_1\gamma_2\ket{H}_{\rm r3} \ket{V}_{\rm s3}
+\delta_2\gamma_1\ket{V}_{\rm r4}\ket{H}_{\rm s4}+\gamma_1\gamma_2\ket{V}_{\rm r4}\ket{V}_{\rm s3}) ], \label{eq:4}
\end{eqnarray}
\end{widetext}
where the numbers in the subscripts  represent the
spatial modes in Fig.\ref{fig:setup}. The polarization rotation now causes photons emerge in port 4, too. In the case where two photons emerge at the port 3, the state of two photons is $\alpha \delta^2_1 \ket{H}_{\rm r3}\ket{H}_{\rm s3}+\beta \gamma^2_2\ket{V}_{\rm r3}\ket{V}_{\rm s3}+\delta_1\gamma_2(\alpha \ket{V}_{\rm r3}\ket{H}_{\rm s3}+\beta \ket{H}_{\rm r3}\ket{V}_{\rm s3})$. The state $\alpha \ket{V}_{\rm r3}\ket{H}_{\rm s3}+\beta \ket{H}_{\rm r3}\ket{V}_{\rm s3}$ is preserved against collective polarization rotations, and it is easy to see that we can post-select the signal state $\alpha \ket{H}+\beta \ket{V}$ by using the same scheme as for collective dephasing shown in Fig. \ref{fig:setupDecode}. This argument shows that we can use conventional optical fibers instead of PMF. When the quantum efficiency of the detector is $\eta$, the overall success probability is given by $(|\delta_1\gamma_2|^2/2)(\eta^2/4)$, where the latter factor represents the loss in the decoding method in Fig. \ref{fig:setupDecode}. In practice, we may wish to have a constant overall success probability, which is independent of fluctuating parameters $\delta_1$ and $\gamma_2$. One easy way to achieve this is to introduce artificial random collective polarization rotations in each channel. This averages the factor $|\delta_1\gamma_2|^2/2$ to $1/8$.

As for the possibility of improving the success probability, we can extract the state $\alpha \ket{H}_{\rm r4}\ket{V}_{\rm s4}+\beta \ket{V}_{\rm r4}\ket{H}_{\rm s4}$ of two photons appearing from port 4 with a similar decoding setup, which doubles the success probability up to 1/4. If a deterministic two-qubit operations can be used, the efficiency will be further improved up to 1/2 by extracting states 
$\alpha \ket{H}_{\rm r4}\ket{H}_{\rm s3}+\beta \ket{H}_{\rm r3}\ket{H}_{\rm s4}$ and 
$\alpha \ket{V}_{\rm r3}\ket{V}_{\rm s4}+\beta \ket{V}_{\rm r4}\ket{V}_{\rm s3}$ from the state (\ref{eq:4}) and decoding them to the signal state. 

We have described a distribution scheme for an arbitrary state of
qubit with an additional qubit, which can be implemented by adding
a single photon in a fixed polarization state. The interesting
feature of our scheme is that Alice only prepares the separable
state of two photons and there is no need to prepare entangled
states. Therefore we can easily apply this scheme to many quantum
communication protocols based on qubit, especially to QKD
protocols. In the following, we show the application  to the
BB84 QKD protocol. To implement BB84 protocol using the present
scheme, Alice randomly chooses one of the four states
$\ket{H}_{\rm s}$, $\ket{V}_{\rm s}$, $\ket{D}_{\rm s}$, and
$\ket{\bar{D}}_{\rm s}$ of the signal photon and sends it along
with the reference photon in $\ket{D}_{\rm r}$. After receiving
and decoding them, Bob measures the decoded photon in the mode Y
in one of the two bases $\{\ket{H}_{\rm Y}, \ket{V}_{\rm Y}\}$ and
$\{\ket{D}_{\rm Y}, \ket{\bar{D}}_{\rm Y}\}$, chosen at random. In
order to confirm that this scheme is equivalent to BB84, let us
assume that Bob only accepts the cases where the mode $X$ is found to
be in $\ket{D}_{\rm X}$. It is not difficult to see that the
projection onto the four states $\ket{H}_{\rm Y}\ket{D}_{\rm X},
\ket{V}_{\rm Y}\ket{D}_{\rm X},\ket{D}_{\rm Y}\ket{D}_{\rm X},
\ket{\bar{D}}_{\rm Y}\ket{D}_{\rm X} $ is respectively equivalent
(up to the normalization) to the projection onto the states
$\ket{h}_{\rm rs},\ket{v}_{\rm rs}, \ket{h}_{\rm rs}+\ket{v}_{\rm
rs},\ket{h}_{\rm rs}-\ket{v}_{\rm rs}$ of the two pulses received
by Bob, where $\ket{h}_{\rm rs}\equiv \ket{V}_{\rm r}\ket{H}_{\rm
s}+\ket{HV}_{\rm r}\ket{\rm vac}_{\rm s}$ and $\ket{v}_{\rm
rs}\equiv \ket{H}_{\rm r}\ket{V}_{\rm s}+\ket{{\rm vac}}_{\rm
r}\ket{HV}_{\rm s}$.
 Let us
introduce a virtual two-qubit system AB by the following relations:
\begin{eqnarray}
&& \sqrt{6}\ket{0}_{\rm A}\ket{0}_{\rm B}\equiv 2\ket{V}_{\rm r}\ket{H}_{\rm s}
+\ket{H}_{\rm r}\ket{H}_{\rm s}+\ket{HV}_{\rm r}\ket{\rm vac}_{\rm s}
\nonumber \\
&& \sqrt{6}\ket{1}_{\rm A}\ket{0}_{\rm B}\equiv 2\ket{H}_{\rm r}\ket{V}_{\rm s}
+\ket{V}_{\rm r}\ket{V}_{\rm s}+\ket{{\rm vac}}_{\rm r}\ket{HV}_{\rm s}
\nonumber \\
&& \sqrt{2}\ket{0}_{\rm A}\ket{1}_{\rm B}\equiv
\ket{H}_{\rm r}\ket{H}_{\rm s}-\ket{HV}_{\rm r} \ket{\rm vac}_{\rm s}
\nonumber \\
&& \sqrt{2}\ket{1}_{\rm A}\ket{1}_{\rm B}\equiv
\ket{V}_{\rm r}\ket{V}_{\rm s}-\ket{{\rm vac}}_{\rm r}\ket{HV}_{\rm s}.
\end{eqnarray}
In view of this, Alice always prepares qubit B in the state
$\sqrt{3}\ket{0}_{\rm B}+\ket{1}_{\rm B}$ and qubit A in one
of the four BB84 states. Bob's measurement can be regarded as first
making sure that the qubit B
is projected onto the state $\sqrt{3}\ket{0}_{\rm B}-\ket{1}_{\rm B}$,
then measuring qubit A as in the BB84 protocol. Therefore,
the whole protocol is equivalent to BB84 except for the efficiency,
and the same security proof holds at least when an ideal single-photon
source is used.

\begin{figure}[tbp]
\begin{center}
 \includegraphics[scale=1]{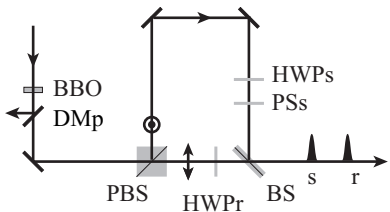}
\end{center}
 \caption {The schematics of generation of signal and reference photons.
 Two photons are collinearly generated by 
 PDC from Type II phase-matched $\beta$-barium borate (BBO) crystal. One is in  $\ket{H}$ and the other is in  $\ket{V}$. They are split into two optical paths by a PBS, then mixed by BS. They are separated by $\Delta t$, which corresponds to the path difference between the short and long path. HWP$_{\rm r}$ rotates the polarization by 45$^\circ$, while HWP$_{\rm s}$ and phase shifter (PS$_{\rm s}$) are used to transform the state $\ket{V}$ into the desired signal state. Two photons in one output port of BS are in the state (\ref{eq:1}).
}
\label{fig:setupSource}
\end{figure}

One of the advantages of the present scheme applied to BB84
protocol is that quantum communication is one-way. This enables us
to avoid the Trojan horse attack which is threatening especially
the Plug-and-play QKD system using two-way quantum communication\cite{Gisin02}
to compensate the fluctuation in optical fiber. Unlike other
schemes \cite{Wang03,Walton03,Boileau04,Boileau04qph} sharing the one-way feature, the present scheme does not
require a preparation of an entangled photon state by Alice, so it
can be easily implemented by various photon sources. 

Finally, we discuss possible relaxations in the requirement for the photon sources used in the present scheme. One of the unavoidable errors caused by the use of imperfect photon sources is  multi-photon emission in the same time bin. For example, if the reference pulse (and/or signal pulse) has two photons, both of them can pass through the path L and register a coincidence detection between the paths X and Y, at the ``correct'' timing. We cannot distinguish this false coincidence from the true ones, even by post-selection. 

In order to make the contribution of these errors sufficiently small, we must use a photon source satisfying the condition $P^{(1,1)}\gg P^{({\rm mul})}$, where $P^{(1,1)}$ is the probability of emitting one photon in the signal and one in the reference, and $P^{({\rm mul})}$ is that of emitting  multiple photons either in the signal or in the reference. 
For a source using parametric down conversion (PDC) shown in Fig. \ref{fig:setupSource}, at least two (almost) independent emissions of photon pairs are necessary for $P^{({\rm mul})}$, and hence satisfies $P^{({\rm mul})}\sim O((P^{(1,1)})^2)$ fulfilling the above condition

On the other hand, in the case of using weak coherent states for both signal and reference, it is difficult to satisfy the above condition for the following reason: Let $\nu$ and $\mu$ be the average photon numbers of the signal and the reference. Since the probability of finding $n$ photons in a coherent state obeys the Poisson statistics, we have $ P^{(1,1)}=e^{-(\nu+\mu)}\nu\mu $ and $ P^{({\rm mul})}\ge e^{-(\nu+\mu)}(\nu^2+\mu^2)/2 $. This shows $P^{(1,1)}\le P^{({\rm mul})}$, so we cannot satisfy the above condition. Of course, if Bob uses two-photon absorber to reduce $P^{({\rm mul})}$, then we can satisfy the above condition. While this indicates an interesting fact that Alice,  in principle, can use completely classical light sources, such schemes are beyond the current technology.

Another possibility is to use a triggered single photon source based on PDC for one pulse \cite{Hong86}, and a  weak coherent state for  the other. As demonstrated in ref. \cite{Rarity97}, we can still observe two-photon interference for this combination. If the average photon number of the coherent state is sufficiently small, the above condition can be met.

In conclusion, we present a  distribution scheme for polarization states of a photon, which is immune to the collective noise. The present scheme does not employ an entangled state encoded in DF subspaces. The quantum state is protected, in principle, with the probability 1/2 under both collective dephasing and depolarization by using two quantum channels,  parity checking, and post-selection. The implementation of the present scheme can be achieved by linear optics and photon detectors. It does not need the stabilization of optical paths with the precision of the order of wavelength. The present scheme is a simple and flexible one that can be used in other quantum communication protocols, besides QKD,  by using various photon sources.

We thank Kiyoshi Tamaki for helpful discussions. This work was supported by 
21st Century COE Program by the Japan Society for the Promotion of Science, 
a MEXT Grant-in-Aid for Young Scientists (B) 17740265, 
 and the National Institute of Information and Communications Technology (NICT).

\end{document}